\documentclass[11pt]{article} 
\usepackage{rldmsubmit,palatino}
\usepackage{graphicx}
\usepackage{amsmath,amssymb}
\usepackage{listings}

\usepackage{algorithm}
\usepackage{algorithmic}
\usepackage{tikz}
\usetikzlibrary{arrows,positioning,shapes,calc}

\title{SCOOP: A Framework for   Proactive Collaboration and Social Continual Learning  through Natural Language Interaction and Causal Reasoning}

\author{
Dimitri Ognibene, Sabrina Patania, Luca Annese, Cansu Koyuturk,  Franca Garzotto, Giuseppe Vizzari  \\
University of Milan-Bicocca\\
Milan, Italy\\
\texttt{dimitri.ognibene@unimib.it} \\
\And
Azzurra Ruggeri \\
TUM School of Social Sciences and Technology \\
Munich, Germany \\
\texttt{a.ruggeri@tum.de} 
\And
Simone Colombani \\
Oversonic Robotics \\
Carate Brianza, Italy \\
}

\begin{document}

\maketitle

\begin{abstract}

Multimodal information-gathering settings, where users collaborate with AI in dynamic environments, are increasingly common. These scenarios involve complex processes with textual and multimodal interaction (e.g., house refurbishment plans) and often require accessing additional structural information (e.g., regulations) via cost-incurring requests. Moreover, AI helpers lack access to users’ true goals, beliefs, and preferences and struggle to integrate diverse information effectively.

We propose a social continual learning framework for causal knowledge acquisition and collaborative decision-making. It focuses on autonomous agents learning through dialogues, question-asking, and interaction in open, partially observable environments. A key component is a natural language oracle that answers the agent’s queries about environmental mechanisms and states, refining causal understanding while balancing exploration (learning) and exploitation (using knowledge).

Evaluation tasks, inspired by developmental psychology, emphasize causal reasoning and question-asking skills, complementing benchmarks by assessing the agent’s ability to identify knowledge gaps, generate meaningful queries, and incrementally update reasoning. The framework also evaluates how the cost of acquiring knowledge is amortized across tasks in the same environment.

We propose two architectures: (1) a system combining Large Language Models (LLMs) with the ReAct framework and question-generation, and (2) an advanced system with a causal world model (symbolic, graph-based, or subsymbolic) for reasoning and decision-making. The latter builds a causal knowledge graph for efficient inference and adaptability under constraints. Challenges include integrating causal reasoning into ReAct and optimizing exploration and question-asking in error-prone scenarios. Beyond applications, this framework models developmental processes combining causal reasoning, question generation, and social learning.

\end{abstract}

\acknowledgements{We acknowledge support from the Volkswagen Foundation for the project \textit{Developing an Artificial Social Childhood (ASC) to improve AI causal reasoning, information gathering and decision making}, Ref.: 9E530.}

\newpage
\section{Introduction}
With the advent of advanced generative AI, tasks involving rich multimodal and natural language information—where users collaborate with AI helpers—are becoming increasingly common. These settings often involve:

\begin{enumerate}       
    \item \textbf{Sequential Collaborative Interactions:} performing the task requires planning and executing of  sequences of interactions and independent by the user or the agent (e.g. collecting different elements and tools), with the need to account for their consequences over multiple steps.
    \item \textbf{Costly Social Information Access}: users and agents must manage limited resources when accessing  information provided by other agents (e.g., mechanisms descriptions, regulations or user preferences or beliefs \cite{Bianco2020}).
        \item \textbf{Multimodal complexity}: combining textual descriptions, visual representations, and structural information (e.g., house refurbishment plans).
     \item \textbf{Dynamic Environments}: evolving tasks with shifting states and requirements (e.g., bureaucratic processes, repair scenarios) \cite{taniguchi2023world}.
\end{enumerate}
The simultaneous presence of these conditions and their interaction motivate our work to formalize and create effective solutions. Moreover, they reflect important aspects of human developmental processes \cite{cangelosi2010integration,ruggeri2016sources}.
\subsection{Example Scenarios}

\textbf{Example 1: Repairing and Shipping Items}

A robot agent assists in repairing and shipping items to various international destinations. Tasks include adhering to specific repair requirements, customs regulations, packaging standards, and climate conditions. The agent collaborates with the user to clarify repair strategies and queries a natural language oracle for regulatory or technical information. During downtime, the agent explores tools and equipment to refine its understanding of repair techniques, improving future performance.

\textbf{Example 2: Compiling Documents for Official Requests}

An agent aids a user in compiling documents for official applications, such as visas or tax submissions. Tasks involve identifying required documents, extracting relevant information, and validating regulations via a natural language oracle. When ambiguities arise, the agent interacts with the user for clarification. In idle periods, it autonomously explores document templates and regulatory archives to improve performance across multiple problem instances.

 \section{Related Work}
\textbf{ReAct Framework:} The ReAct (Reason + Act) framework \cite{yao2023} integrates reasoning and acting capabilities into LLMs, enabling contextual decision-making and blending question generation with action-taking in collaborative environments. 

\textbf{Causal Reasoning in AI:} Techniques like causal discovery \cite{peters2017elements} and inference frameworks such as DoWhy \cite{sharma2020dowhy} provide essential tools for reasoning in dynamic, partially observable environments.

\textbf{Integrating Causal Graphs with LLMs:} Recent research integrates causal graphs with LLMs to enhance reasoning and decision-making. Embedding causal reasoning into LLM workflows enables systems to interpret environments, predict outcomes, and optimize actions. Frameworks combining LLMs with causal world models \cite{gkountouras2024languageagentsmeetcausality} and studies on automating causal discovery \cite{long2024largelanguagemodelsbuild} highlight the potential of causal representation learning for dynamically constructing world models.

\textbf{LLMs and Planning Systems:} Combining LLMs with planning systems, such as symbolic planners or reinforcement learning frameworks, supports structured decision-making. Hybrid approaches integrate LLMs for language understanding and planners for task execution, excelling in complex scenarios like multi-agent collaboration and robotics \cite{colombani2024one,kambhampati2024llms}.

\textbf{Causal Discovery and Question Generation:} Integrating LLMs with causal discovery tools enhances reasoning in dynamic environments. For instance, PyWhy-LLM supports causal analysis \cite{kiciman2023causal}, while DoWhy-GCM facilitates inference in graphical causal models \cite{blobaum2024dowhy}. These tools enable LLMs to refine causal graphs and resolve ambiguities, closing the loop between knowledge acquisition and action.

\textbf{Symbolic and Subsymbolic Integration:} Hybrid architectures bridge symbolic causal reasoning with the subsymbolic capabilities of LLMs, combining graph-based reasoning with unstructured data processing for robust decision-making agents \cite{hitzler2022neuro, Ibrahim2024}.

\textbf{Active Information Gathering:} Research on active learning and information gain \cite{patania2024large, bertolazzi2023chatgpt, friston2015active, ognibene2019proactive, masiero2024search} informs strategies for querying to maximize utility while balancing exploration and exploitation in interactive systems.

\textbf{Human-in-the-Loop Systems:} Human feedback integration highlights the value of interactive learning, enabling agents to dynamically query users and adapt to preferences \cite{amershi2014power}.

\textbf{Developmental Psychology Insights:} Insights from children’s causal learning \cite{gopnik2004theory, legare2014curiosity} inspire benchmarks to evaluate AI agents’ abilities in question generation and causal inference.

By synthesizing these elements, our framework advances adaptive, interactive AI systems at the intersection of causal reasoning, social learning, and multimodal interaction.

\section{The SCOOP framework: Social Continual Object-Oriented POMDP }
\paragraph{Formal Framework.}
We formalize our rich interactive setting as an \emph{object-oriented} partially observable Markov decision process (OO-POMDP), extended to incorporate a \emph{lingusistic world descriptor} generating natural language observations  (e.g. the initial task description in natural language), multiple problem instances, a \emph{user} with problem-specific objectives, and a \emph{natural language oracle} that provides causal information. Let 
\[
\mathcal{D} = \bigl(\mathcal{T}_O, \mathcal{F}_O, \mathcal{R}_O, \mathcal{W}_O \bigr)
\]
denote a \emph{domain specification}, where: 
\begin{itemize}
    \item \(\mathcal{T}_O\) is a set of object types (e.g., boxes, tools, containers);
    \item \(\mathcal{F}_O\) is the set of allowed features or predicates relevant to the domain (e.g., ``contains(x,y)'' or ``isOpenable(x)'');
    \item \(\mathcal{R}_O\) is an evolving set of \emph{causal rules} capturing how these features and object types interact (e.g., whether opening a container allows access to its contents), some of which may be unknown or partially specified;
    \item \(\mathcal{W}_O\) is the family of possible world configurations consistent with \(\mathcal{T}_O, \mathcal{F}_O,\) and \(\mathcal{R}_O\).
\end{itemize}

A \emph{problem instance} \(\theta\) refines \(\mathcal{D}\) with a concrete set of objects, an initial state, and a \emph{user objective} (defined by a problem-specific reward function \(r_\theta\) that encodes the user’s goals). The \emph{helper agent} interacts with both the user,  making questions to clarify his goals or preferences, presenting results and suggestions, or other interactive actions, and a \emph{natural language oracle} (to acquire missing causal rules or environment states) across potentially many problem instances \(\{\theta_i\}\). The oracle responds in multiple formats:
\begin{enumerate}
    \item \emph{Language-based descriptions} of environment dynamics, such as “box A must be opened before retrieving item B”;
    \item \emph{Formal causal chunks}, where the oracle may directly provide rules or graphs parts (e.g., “node \(\texttt{Open(Box)}\) causes \(\texttt{Accessible(Item)}\)”),
    \item \emph{Observation-like feedback}, akin to sensor readings or state confirmations.
\end{enumerate}

The overall action space \(\mathcal{A}=\mathcal{A}^a \cup \mathcal{A}^u\)  composed by  the agent’s \emph{action space} by \(\mathcal{A}^a\) and  the user’s \emph{action space} by \(\mathcal{A}^u\). 
We denote the agent’s \emph{action space} by \(\mathcal{A}^a = \mathcal{A}^a_\mathrm{act}\cup \mathcal{A}^a_\mathrm{query}\), where \(\mathcal{A}_\mathrm{act}\) includes environment-oriented actions (e.g., open, pick, place) and \(\mathcal{A}_\mathrm{query}\) includes queries to the oracle or the user. 
The agent also observes the user’s feedback or clarifications regarding the task objectives and environment states. We denote the user’s \emph{action space} by \(\mathcal{A}^u = \mathcal{A}^u_\mathrm{act}\cup \mathcal{A}^u_\mathrm{query}\), where \(\mathcal{A^u}_\mathrm{act}\) includes environment-oriented actions (e.g., open, pick, place) and \(\mathcal{A}_\mathrm{query}\) includes queries to the agent. Formally, each problem instance is modeled as:
\[
(\mathcal{S}_\theta, \mathcal{A}, \Omega_\theta, \mathcal{T}_\theta, \mathcal{O}_\theta, r^u_\theta,r^a_\theta, \gamma, \beta),
\]
mirroring a POMDP with the following modifications:
\begin{itemize}
    \item \(\mathcal{S}_\theta\) embeds the object-based states from \(\theta\) and any partial knowledge of the causal rules \(\mathcal{R}_O\);
    \item \(\Omega_\theta\) is the space of possible observations, spanning both \emph{environmental signals} (e.g., sensor readings, gripper state, environment map chunk, etc) and \emph{language-based} responses from the user or oracle;    
    \item \(r^u_\theta\) encodes the \emph{user’s objectives} for instance \(\theta\), which the helper agent aims to optimize;
    \item \(r^a_\theta\) encodes the \emph{helper agent's action costs} for instance \(\theta\), which the helper agent aims to optimize;
    \item \(\beta(\cdot)\) defines the \emph{cost of querying} (time, resources, or complexity)  the oracle to obtain new causal information or 
 the user about current objective and preferences, i.e. \(r^u_\theta\).
\end{itemize}

Crucially, the agent can \emph{explore} the domain outside active tasks to refine \(\mathcal{R}_O\) (e.g., by performing experiments or asking domain-level questions). Any information gleaned is \emph{amortized} across future tasks \(\theta_j\). This design enables \textbf{continual learning} of domain mechanics: as the agent accumulates causal knowledge (e.g., “a certain box can contain items of type \(T\)”), it improves performance in subsequent problem instances. More formally this is obtained assuming that at each instance \(\theta_j\) the specific instantiation of \(\mathcal{R}_O\),   \(r^u_\theta\) and  \(r^a_\theta\) are extracted from the same distribution.

Ultimately, the helper agent objective functions is:
\(\sum_\theta \sum_t^{T(\theta)}\gamma^{t+T(-\theta)}(r^u_\theta(s_t,a^u_t)+r^a_\theta(s_t,a^a_t)+\beta(a^a_t))\).
Balancing \emph{exploration} (question-asking, active experimentation) and \emph{exploitation} (leveraging current knowledge to solve tasks efficiently) is thus a central challenge in this social continual learning framework.

\section{Developmental Psychology-Inspired Tasks for Evaluating Causal Reasoning and Question-Making}

Drawing from developmental psychology, we design tasks to evaluate causal reasoning and question-making skills in collaborative AI systems. Inspired by children’s learning behaviors, these tasks assess the agent's ability to:

\textit{Explore-Exploit Tradeoff}: Balance between directed exploration and utilizing known information to reduce uncertainty and achieve goals efficiently \cite{meder2021development}.
\textit{"Why" and "What If" Questions}: Formulate meaningful hypotheses and evaluate counterfactual scenarios to refine causal models \cite{walker2020asking}.
\textit{Epistemic Question Formulation}: Construct precise, goal-directed queries to address knowledge gaps efficiently \cite{ronfard2018question}.
\textit{Causal Inference and Learning}:
Engage AI in tasks where it observes incomplete sequences of events and must infer causal relationships. For example, after observing that certain components drive a machine, the AI predicts outcomes without direct trial-and-error \cite{gopnik2001causal,shavlik2022contributions}.
\textit{Generating Hypotheses from Confounded Evidence}: Assesses AI's ability to generate interventions that are informative in resolving the structure of ambiguous causal system \cite{gweon2008stretching}.
These tasks provide a multidimensional framework to benchmark AI systems, focusing on cognitive, linguistic, and social reasoning capabilities essential for dynamic, real-world collaboration.

\section{Reference architectures}
\subsection{Base: Oracle-Aided ReAct}
The base architecture simply extends the ReAct framework introducing actions to ask  state and the user about their preference and objectives and the oracle about environment mechanics. However, as noted in the literature, complex planning \cite{katz2024thought} and usage of declarative knowledge for  decision making and execution \cite{arabi2024habit} appear difficult for vanilla LLMs.

\subsection{Advanced: ReAct Framework with Oracle-Aided Causal Reasoning}

Building on the base architecture, the advanced ReAct (Reason + Act) \cite{yao2023} framework introduces information-gathering actions and extends its functionality with a specialized action, \texttt{CausalRefinementAndAction}. This action is invoked by the Large Language Model (LLM) when complex reasoning tasks are required, specifically for:

\begin{itemize}
    \item \textbf{Refining or updating} knowledge about user needs and the world's mechanisms and states (causal model), or
    \item \textbf{Planning and executing} steps to achieve a specified goal.
\end{itemize}

\texttt{CausalRefinementAndAction} integrates iterative causal knowledge management, utilizing both external oracle support (e.g., a domain expert or automated simulator) and established causal inference libraries such as \textit{causal-learn}, \textit{DoWhy}, and \textit{Tetrad} \cite{zheng2024causal,sharma2020dowhy,tetrad}. Given a user’s prompt, current goal, and contextual information, the LLM initially maps relevant knowledge to a causal graph, which may be incomplete.

The agent estimates the expected value and cost of potential actions to refine its knowledge, using metrics such as Value of Information (VoI) or robust optimization criteria. Refinement actions include querying the user about preferences and goals, asking the oracle about specific causal links or effect sizes, or performing interventions. If the refinement is deemed beneficial (i.e., cost is below a threshold), the suggested strategy is executed. Based on responses from the user or oracle, or results of interventions, causal inference libraries update the graph and determine whether additional refinements are necessary.

Once the causal graph is sufficiently refined, the ReAct agent invokes planning routines—using libraries such as \textit{PyCID} or a robust Markov Decision Process (MDP) solver—to derive policies or action sequences that maximize the likelihood of achieving the user’s goals under the current causal knowledge. This combination of LLM-driven reasoning, causal knowledge management, and decision-making enable advanced reasoning and information-gathering capabilities are activated when necessary while maintaining the flexibility to handle diverse scenarios typical of LLMs. 
\bibliographystyle{plain}
\bibliography{bib-short}





\end{document}